\documentclass[12pt]{article}
\usepackage[cp1251]{inputenc} 
\usepackage[english, russian]{babel} 
\usepackage{amsmath, latexsym,  amssymb, bm,  graphics, eucal,
amsfonts, array}
\usepackage[dvips]{graphicx}
\makeatletter
\renewcommand{\@biblabel}[1]{#1.\hfill}

\newcommand{\const}{\mathop{\rm const\, }}

\makeatother
\renewcommand{\baselinestretch}{1.1}
\newcommand{\be}{\begin{center}}
\newcommand{\ee}{\end{center}}

\begin{document}

\newcommand{\mc}[1]{\mathcal{#1}}
\newcommand{\E}{\mc{E}}
\thispagestyle{empty}
\large

\renewcommand{\abstractname}{\,Abstract}
\renewcommand{\refname}{\begin{center} REFERENCES\end{center}}

 \begin{center}
\bf Longitudinal dielectric permeability in quantum non-degenerate
and maxwellian collisional plasma with constant collision frequency
\end{center}\medskip

\begin{center}
  \bf A. V. Latyshev\footnote{$avlatyshev@mail.ru$} and
  A. A. Yushkanov\footnote{$yushkanov@inbox.ru$}
\end{center}\medskip

\begin{center}
{\it Faculty of Physics and Mathematics,\\ Moscow State Regional
University, 105005,\\ Moscow, Radio str., 10--A}
\end{center}\medskip

\begin{abstract}
The formula for dielectric function of non-degenerate and
maxwellian collisional plasmas is transformed to the form,
convenient for research.
Graphic comparison of longitudinal dielectric functions
of quantum and classical non-degenerate collisional plasmas is
made.


\end{abstract}

\begin{center}
{\bf Introduction}
\end{center}

In the known work of Mermin \cite{Mermin} by means of the analysis
of nonequilib\-rium density matrix in $\tau $--approximation
has been obtained expression for longitu\-di\-nal dielectric permeability of
quantum collisional plasmas for case of constant collision frequency of plasmas
particles.

Earlier in the work of Klimontovich and Silin \cite{Klim} and after that in
the work of Lindhard \cite{Lin} has been obtained
expression for longitudinal and transverse dielectric permeability of quantum
collisionless plasmas.

In work \cite{Mermin} actually without deducing it has been announced
the general expression of dielectric function of quantum
collisional plasmas with constant frequency of collisions.

In work \cite{Lat2012} the general expression
for dielectric function of quantum collisio\-nal plasmas with
the variable frequency of collisions depending on the wave
vector  has been deduced.

Then in work \cite{Lat2012a} the detailed deducing dielectric
functions for quantum plasma with constant frequency of collisions
has been given.

In the present work the case of non-degenerate and maxwellian
quantum plasma with any degree of degeneration (a case of the any
temperature)  is considered.

The general formula for dielectric functions from
\cite {Mermin}, \cite {Lat2012} and \cite {Lat2012a}
it becomes simpler and transformed to the convenient form
for calculations and contains one quadrature.

Deducing of dielectric function of classical
non-degenerate collisional plas\-mas is presented.
It is given graphic comparison of the real and imaginary parts of
dielectric function of quantum and classical plasma.

As classical non-degenerate collisional plasma we will be
understand the plasma described by the kinetic Vlasov---Boltzmann equation
with integral of collisions from phase space.

As quantum non-degenerate collisional plasma we will be
understand the plasma described by the kinetic Schr\"{o}dinger---Boltzmann
equation with integral of collisions from momentum space.

Quantum plasma is studied extremely intensively. Among
the big number of works we will note only some of them
\cite{Kliewer}--\cite{Ropke}.

\begin{center}\bf
1.  Dielectric permeability of quantum non-degenerate plasmas
\end{center}

We take the formula for longitudinal dielectric function of quantum
non-degenerate collisional plasmas
$$
\varepsilon_l({\bf k},\omega,\nu)=1+\dfrac{4\pi e^2}{k^2}
\dfrac{(\omega+i \nu)B({\bf k},\omega+i \nu)B({\bf k},0)}
{\omega B({\bf k},0)+i \nu B({\bf k},\omega+i \nu)}.
\eqno{(1.1)}
$$

In the formula (1.1) following designations are accepted
$$
B({\bf k},\omega+i \nu)=\int \dfrac{d{\bf p}}{4\pi^3}\dfrac{f_{{\bf p+k}/2}-
f_{{\bf p-k}/2}}{\E_{{\bf p-k}/2}-\E_{{\bf p+k}/2}+\hbar(\omega+i \nu)},
\eqno{(1.2)}
$$

$$
B({\bf k},0)=\int \dfrac{d{\bf p}}{4\pi^3}\dfrac{f_{{\bf p+k}/2}-f_{{\bf p-k}/2}}
{\E_{{\bf p-k}/2}-\E_{{\bf p+k}/2}},
$$\medskip
$\bf k$ is the dimensional wave vector,
$$
\E_{{\bf p\pm k}/2}=\dfrac{\hbar^2}{2m}
\Big({\bf p}\pm \dfrac{{\bf k}}{2}\Big)^2,\qquad
f_{{\bf p\pm k}/2}=\dfrac{1}{1+\exp\Big(\dfrac{\E_{{\bf p\pm k}/2}}
{k_BT}-\alpha\Big)},
$$
$\alpha=\mu/(k_BT)$ is the normalized (dimensionless) chemical
potential, $k_B$ is the Boltzmann constant, $T$ is the temrerature,
$f_{{\bf p\pm k}/2}$ is the Fermi---Dirac distribution function, $e$
is the electron charge, $\nu$ is the effective collision
frequency of electrons with plasma particles, $\omega$ is the
frequency of oscillations of an electromagnetic field.

Let's transform the formula (1.1) to the form convenient for research.
We will enter thermal velocity of electrons $v_T=1/\sqrt{\beta}$,
$\beta=m/(2k_BT)$. Clearly, that $k_BT=\dfrac{mv_T^2}{2}=\E_T$
is thermal electrons energy.

We introduce the nondimensional vector ${\bf P}=\dfrac{{\bf p}}{k_T}$,
where $k_T=\dfrac{{\bf p_T}}{\hbar}=\dfrac{mv_T}{\hbar}$ is the themal
wave number, ${\bf p}_T$ is the thermal momentum of electrons.

Let's find  difference of energies
$$
\E_{{\bf p-k}/2}-\E_{{\bf p+k}/2}=-2\E_TP_xq,\qquad {\bf q}=q(1,0,0),\qquad
{\bf q}=\dfrac{{\bf k}}{k_T},
$$
${\bf q}$ is the nondimensional wave number.

The denominator from expression (1.2) is equal
$$
\E_{{\bf p-k}/2}-\E_{{\bf p+k}/2}+\hbar(\omega+i \nu)=
-\dfrac{\hbar^2k_T^2}{m}q\Big(P_x-\dfrac{z}{q}\Big)=
-2\E_Tq(P_x-\dfrac{z}{q}).
$$

Here dimensionless frequencies are entered
$$
z=x+iy=\dfrac{\omega+i \nu}{k_Tv_T},\qquad x=\dfrac{\omega}{k_Tv_T},
\qquad y=\dfrac{\nu}{k_Tv_T}.
$$

Now the integral (1.2) is equal
$$
B({\bf k},\omega+i \nu)=-\dfrac{k_T^3}{2\E_Tq}\int \dfrac{d^3P}{4\pi^3}
\dfrac{f_{{\bf p+q}/2}-f_{{\bf p-q}/2}}{P_x-z/q}.
\eqno{(1.3)}
$$

Here
$$
f_{{\bf p\pm q}/2}=
\dfrac{1}{1+\exp\Big[\Big({\bf P}\pm\dfrac{{\bf q}}{2}\Big)^2-\alpha\Big]}
\equiv f_{{\bf P\pm q}/2}.
$$

Let's designate
$$
B({\bf q},z)=\int \dfrac{d^3P}{4\pi^3}
\dfrac{f_{{\bf P+q}/2}-f_{{\bf P-q}/2}}{P_x-z/q}.
\eqno{(1.4)}
$$

Then expression (1.3) will be rewritten in the form
$$
B({\bf k},\omega + i \nu)=-\dfrac{k_T^3}{2\E_Tq}B({\bf q},z).
$$

Let's consider integral (1.4). We will present this integral in the form
of difference of integrals. In each of integrals it is realizable the obvious
linear replacement of variables. As result we receive
$$
B({\bf q},z)=\int \dfrac{d^3P}{4\pi^3}\dfrac{f_0(P,\alpha)}{P_x-z/q-q/2}-
\int \dfrac{d^3P}{4\pi^3}\dfrac{f_0(P,\alpha)}{P_x-z/q+q/2},
\eqno{(1.5)}
$$
where
$$
f_0(P,\alpha)=\dfrac{1}{1+e^{P^2-\alpha}}.
$$
According to (1.5) we receive
$$
B({\bf k},\omega+i \nu)=-\dfrac{k_T^3}{2\E_T}b(q,z),
\eqno{(1.6)}
$$
where
$$
b(q,z)=
\int \dfrac{d^3P}{4\pi^3}\dfrac{f_0(P,\alpha)}{(P_x-z/q)^2-(q/2)^2}.
$$

Now expression (1.1) for dielectric function will be transformed
to the form
$$
\varepsilon_l(x,y,q)=1-\dfrac{4\pi e^2}{q^2k_T^2}\dfrac{k_T^3}{2\E_T}
\dfrac{(x+iy)b(q,z)b(q,0)}{xb(q,0)+iyb(q,z)}.
\eqno{(1.7)}
$$

It is easy to find, that numerical density of particles of plasma
(its concentration) in an equilibrium condition  is equal
$$
N=\int f_0(P,\alpha)\dfrac{2m^3d^3v}{(2\pi \hbar)^3}=
\dfrac{2m^3v_T^3}{8\pi^3\hbar^3}\int f_0(P,\alpha)d^3P=
\dfrac{k_T^3}{\pi^2}f_2(\alpha),
\eqno{(1.8)}
$$
where
$$
f_2(\alpha)=\int\limits_{0}^{\infty}f_0(P,\alpha)P^2dP.
$$

Let's calculate internal integral on $dP_ydP_z $ in expression for
$b(q,z)$. We have
$$
\int\limits_{-\infty}^{\infty}\int\limits_{-\infty}^{\infty}
\dfrac{dP_ydP_z}{1+\exp(P_x^2+P_y^2+P_z^2-\alpha)}=\pi\ln(1+e^{-P_x^2}).
$$

Hence, the integral $b(q,z)$ is expressed through the
one-dimensional integral
$$
b(q,z)=\dfrac{1}{4\pi^2}l(q,z),
\eqno{(1.9)}
$$
where
$$
l(q,z)=\int\limits_{-\infty}^{\infty}\dfrac{\ln(1+e^{\alpha-\mu^2})d\mu}
{(\mu-z/q)^2-(q/2)^2}.
$$

Longitudinal dielectric function (1.7) according to (1.8) and (1.9)
we rewrite  in the form
$$
\varepsilon_l(x,y,q)=1-\dfrac{x_p^2}{4q^2f_2(\alpha)}\dfrac{(x+iy)l(q,z)l(q,0)}
{xl(q,0)+iyl(q,z)}.
\eqno{(1.10)}
$$

In (1.10) $x_p$ is the nondimensional plasma frequency,
$$
x_p=\dfrac{\omega_p}{k_Tv_T},
$$
$\omega_p$ is the dimensional plasma (Langmuir) frequency,
$$
\omega_p=\dfrac{4\pi e^2N}{m}.
$$

\begin{center}\bf
2.  Dielectric permeability of classical non-degenerate plasmas
\end{center}

We take the kinetic Vlasov---Boltzmann equation for collisional
Fermi---Dirac plasmas with arbitrary temperature
$$
\dfrac{\partial f}{\partial t}+{\bf v}\dfrac{\partial f}{\partial {\bf r}}+
e{\bf E}({\bf r},t)\dfrac{\partial f}{\partial {\bf p}}=\nu [f_{eq}-f].
\eqno{(2.1)}
$$

Here $f_{eq}$ is the local equilibrium distribution electrons function
of Fermi---Dirac (local Fermian)
$$
f_{eq}=\dfrac{1}{1+\exp\Big(\dfrac{mv^2}{2k_BT}-
\dfrac{\mu({\bf r})}{k_BT}\Big)},
\eqno{(2.2)}
$$
$k_B$ is the Boltzmann constant, $T$ is the plasma temperature, $\nu$
is the electron collisional frequency with plasma particles, ${\bf p}=m{\bf v}$
is the electron momentum, $e$ is the electron charge, $\mu({\bf r})$
is the plasma chemical potential.
We present the chemical potential in linear approximation as
$$
\mu({\bf r})=\mu+\delta \mu({\bf r}), \qquad \mu=\const.
$$

We introduce the nondimensional parameters
$$
t_1=\nu t,\qquad {\bf P}=\dfrac{{\bf v}}{v_T}=\dfrac{{\bf p}}{p_T},
\qquad v_T=\dfrac{1}{\sqrt{\beta}},
$$
$$
{\bf r}_1=\dfrac{{\bf r}}{l_T},\qquad l_T=v_T\tau,\qquad \tau=\dfrac{1}{\nu},
\qquad
\alpha({\bf r})=\dfrac{\mu({\bf r})}{k_BT}.
$$

Let's rewrite equation (2.1) in the following form
$$
\dfrac{\partial f}{\partial t_1}+{\bf P}\dfrac{\partial f}{\partial {\bf r}_1}+
\dfrac{e\tau}{mv_T}{\bf E}({\bf r}_1,t_1)=f_{eq}({\bf P},\alpha({\bf r}_1))-
f({\bf r}_1,{\bf P},t_1).
\eqno{(2.2)}
$$
Here
$$
f_{eq}({\bf P},\alpha({\bf r}_1))=\dfrac{1}{1+\exp(P^2-\alpha({\bf r}_1))},\quad
\alpha({\bf r}_1)=\alpha+\delta \alpha.
$$

We find the distribution function in the form
$$
f=f_0(P,\alpha)+g(P,\alpha)h({\bf r}_1,{\bf P},t_1),
\eqno{(2.3)}
$$
where
$$
f_0(P,\alpha)=\dfrac{1}{1+e^{P^2-\alpha}},\qquad
g(P,\alpha)=\dfrac{e^{P^2-\alpha}}{(1+e^{P^2-\alpha})^2}.
$$.

Let's linearize the equilibrium function $f_{eq}$:
$$
f_{eq}=f_0(P,\alpha)+g(P,\alpha)\delta \alpha.
$$
In linear approximation we have
$$
{\bf E}\dfrac{\partial f}{\partial {\bf p}}={\bf E}\dfrac{\partial f_0}
{\partial {\bf P}}\dfrac{1}{p_T}=-{\bf E}({\bf r}_1,t_1)g(P,\alpha)
\dfrac{2{\bf P}}{p_T}.
$$
Let's sustitute (2.3) into (2.2) and then we obtain
$$
\dfrac{\partial h}{\partial t_1}+{\bf P}\dfrac{\partial h}{\partial {\bf r}_1}
+h({\bf r}_1,{\bf P},t_1)=\dfrac{2e\tau {\bf P}}{p_T}{\bf E}({\bf r}_1,t_1)+
\delta \alpha.
\eqno{(2.4)}
$$

Let's consider the law of preservation of number of particles
$$
\int   (f-f_{eq})d\Omega=0,\qquad d\Omega=\dfrac{2m^3d^3v}{(2\pi\hbar)^3}.
$$
From this law we obtain that
$$
\delta\alpha=\dfrac{\displaystyle
\int g(P,\alpha)h({\bf r}_1,{\bf P},t_1)d^3P}{\displaystyle
g(P,\alpha)d^3P}=\dfrac{1}{2\pi f_0(\alpha)}
\int g(P,\alpha)h({\bf r}_1,{\bf P},t_1)d^3P.
$$
Here
$$
f_0(\alpha)=\int\limits_{0}^{\infty}f_0(P,\alpha)dP.
$$
This equality becomes simpler
$$
\delta\alpha=\dfrac{1}{2\pi f_0(\alpha)}\int\limits_{-\infty}^{\infty}
f_0(\mu',\alpha)h({\bf r}_1,\mu',t_1)d\mu'.
$$

Let's consider further that
$$
h({\bf r}_1,{\bf P},t_1)=h(x_1,P_x,t_1),\qquad
{\bf E}({\bf r}_1,t_1)=e^{i(k_1x_1-\omega_1t_1)}(1,0,0).
$$
We notice that $k_1x_1-\omega_1t_1=kx-\omega t$, because $k_1=kl_T$,
$x_1=x/l_T$, $\omega_1=\omega\tau$, $t_1=\nu t=t/\tau$.

Thus, for function $h$ the following kinetic equation  is received
$$
\dfrac{\partial h}{\partial t_1}+P_x\dfrac{\partial h}{\partial x_1}+
h(x_1,P_x,t_1)=$$$$=\dfrac{2e\tau}{p_T}P_xE_x(x_1,t_1)+
\int\limits_{-\infty}^{\infty}K(\mu',\alpha)h(x_1,\mu',t_1)d\mu'.
$$
Here
$$
K(\mu,\alpha)=\dfrac{f_0(\mu',\alpha)}{2f_0(\alpha)}.
$$

For function $h$ we will search in the form
$h(x_1,\mu,t_1)=\psi(\mu)e^{k_1x_1-\omega_1t_1}$. From equation
(2.4) we find that
$$
\psi(\mu)=\dfrac{e^{-i(k_1x_1-\omega_1t_1)}\delta\alpha+(el_T/\E_T)\mu}
{1-i\omega\tau+ik_1\mu},\qquad \mu=P_x.
\eqno{(2.6)}
$$

Substituting (2.6) in (2.5), we receive:
$$
e^{-i(k_1x_1-\omega_1t_1)}\delta\alpha=\dfrac{el_T}{\E_T}
\dfrac{B_1(k_1,\omega_1)}{1-B_0(k_1,\omega_1)}.
$$

Here
$$
B_1(k_1,\omega_1)=\dfrac{1}{2f_0(\alpha)}
\int\limits_{-\infty}^{\infty}\dfrac{\mu f_0(\mu,\alpha)d\mu}
{1-i\omega_1+ik_1\mu},
$$
$$
B_0(k_1,\omega_1)=\dfrac{1}{2f_0(\alpha)}
\int\limits_{-\infty}^{\infty}\dfrac{f_0(\mu,\alpha)d\mu}
{1-i\omega_1+ik_1\mu}.
$$
It means that according to (2.6) the function $ \psi $ is constructed
$$
\psi(\mu)=\dfrac{el_T}{\E_T}\dfrac{B_1/(1-B_0)+\mu}
{1-i\omega\tau+ik_1\mu},\qquad \mu=P_x.
\eqno{(2.7)}
$$

From definition of density of a current follows, that
$$
{\bf j}=\sigma_l e^{i({\bf kr}-\omega t)}=e\int {\bf v}fd\Omega=
e\int v_x e^{i({\bf kr}-\omega t)} g(P,\alpha)
\psi(\mu)d\Omega.
$$

From here for electrical conductivity we receive
$$
\sigma_l=e\int v_x g(P,\alpha)\psi(\mu)d\Omega.
$$
Let's substitute (2.7) in this equality and we will receive expression for
longitudinal conductivity
$$
\dfrac{\sigma_l}{\sigma_0}=\dfrac{1}{2f_2(\alpha)}\int\limits_{-\infty}^{\infty}
\dfrac{\mu^2+\mu B_1/(1-B_0)}{1-i\omega_1+ik_1\mu}f_0(\mu,\alpha)d\mu=
$$
$$
=\dfrac{1}{2f_2(\alpha)}\Bigg[\int\limits_{-\infty}^{\infty}
\dfrac{\mu^2f_0(\mu,\alpha)d\mu}{1-i\omega_1+ik_1\mu}+\dfrac{B_1}{1-B_0}
\int\limits_{-\infty}^{\infty}
\dfrac{\mu f_0(\mu,\alpha)d\mu}{1-i\omega_1+ik_1\mu}\Bigg],
$$
or
$$
\dfrac{\sigma_l}{\sigma_0}=\dfrac{f_0(\alpha)}{f_2(\alpha)}\Big[
B_2+\dfrac{B_1^2}{1-B_0}\Big].
\eqno{(2.8)}
$$
We notice that
$$
B_1=\dfrac{1}{ik_1}-\dfrac{1-i\omega\tau}{ik_1}B_0,\qquad
B_2=-\dfrac{1-i\omega\tau}{ik_1}B_1.
$$
According to (2.8)
$$
\dfrac{\sigma_l}{\sigma_0}=\dfrac{f_0(\alpha)}{f_2(\alpha)}
\dfrac{(\omega/kv_T)B_1}{1-B_0}=-i \dfrac{xy}{q^2}\dfrac{f_0(\alpha)}
{f_2(\alpha)}\cdot\dfrac{1+(z/q)b(z/q)}{1+(iy/q)b(z/q)}.
\eqno{(2.9)}
$$
Here
$$
b(z/q)=\dfrac{1}{2f_0(\alpha)}\int\limits_{-\infty}^{\infty}
\dfrac{f_0(\mu,\alpha)d\mu}{\mu-z/q}.
$$
On the basis of (2.9) we receive expression for the longitudinal
dielectric function of classical non-degenerate plasmas
$$
\varepsilon_l=1+\dfrac{x_p^2}{q^2}\dfrac{f_0(\alpha)}{f_2(\alpha)}
\cdot\dfrac{1+(z/q)b(z/q)}{1+(iy/q)b(z/q)}.
\eqno{(2.10)}
$$

\begin{center}\bf
  3. Maxwell quantum and classical plasmas
\end{center}

Passing to the limit at $ \alpha\to-\infty $ in the formula (1.10),
we receive expression for longitudinal dielectric function
of quantum Maxwell collisional plasmas
$$
\varepsilon_l(q,x,y)=1-\dfrac{x_p}{q^2}\dfrac{(x+iy)l_0(q,z)l_0(q,0)}
{xl_0(q,0)+iy l_0(q,z)}.
\eqno{(3.1)}
$$

In formula (3.1) the designation is accepted
$$
l_0(q,z)=\dfrac{1}{\sqrt{\pi}}\int\limits_{-\infty}^{\infty}
\dfrac{e^{-\mu^2}d\mu}{(\mu-z/q)^2-(q/2)^2}.
$$

Passing to the limit at $ \alpha\to-\infty $ in the formula (2.10),
we receive expression for longitudinal dielectric function
of classical Maxwell collisional plasmas
$$
\varepsilon_l(q,x,y)=1+\dfrac{2x_p}{q^2}\dfrac{1+(z/q)t(z/q)}
{1+(iy/q)t(z/q)}.
\eqno{(3.2)}
$$

In formula (3.2) the designation is accepted
$$
t(z/q)=\dfrac{1}{\sqrt{\pi}}\int\limits_{-\infty}^{\infty}
\dfrac{e^{-\mu^2}d\mu}{\mu-z/q}.
$$

\begin{center}
\bf  4. Comparison of quantum and classical plasma
\end{center}

Curves $1$ and $2$ on figs. 1--8 correspond to quantum and classical plasmas.
Dimensionless chemical potential  on figs. 1--8 equals to zero: $\alpha=0$.
Dimensionless plasma frequency equals to unit: $x_p=1$.

\begin{figure}[ht]\center
\includegraphics[width=16.0cm, height=10cm]{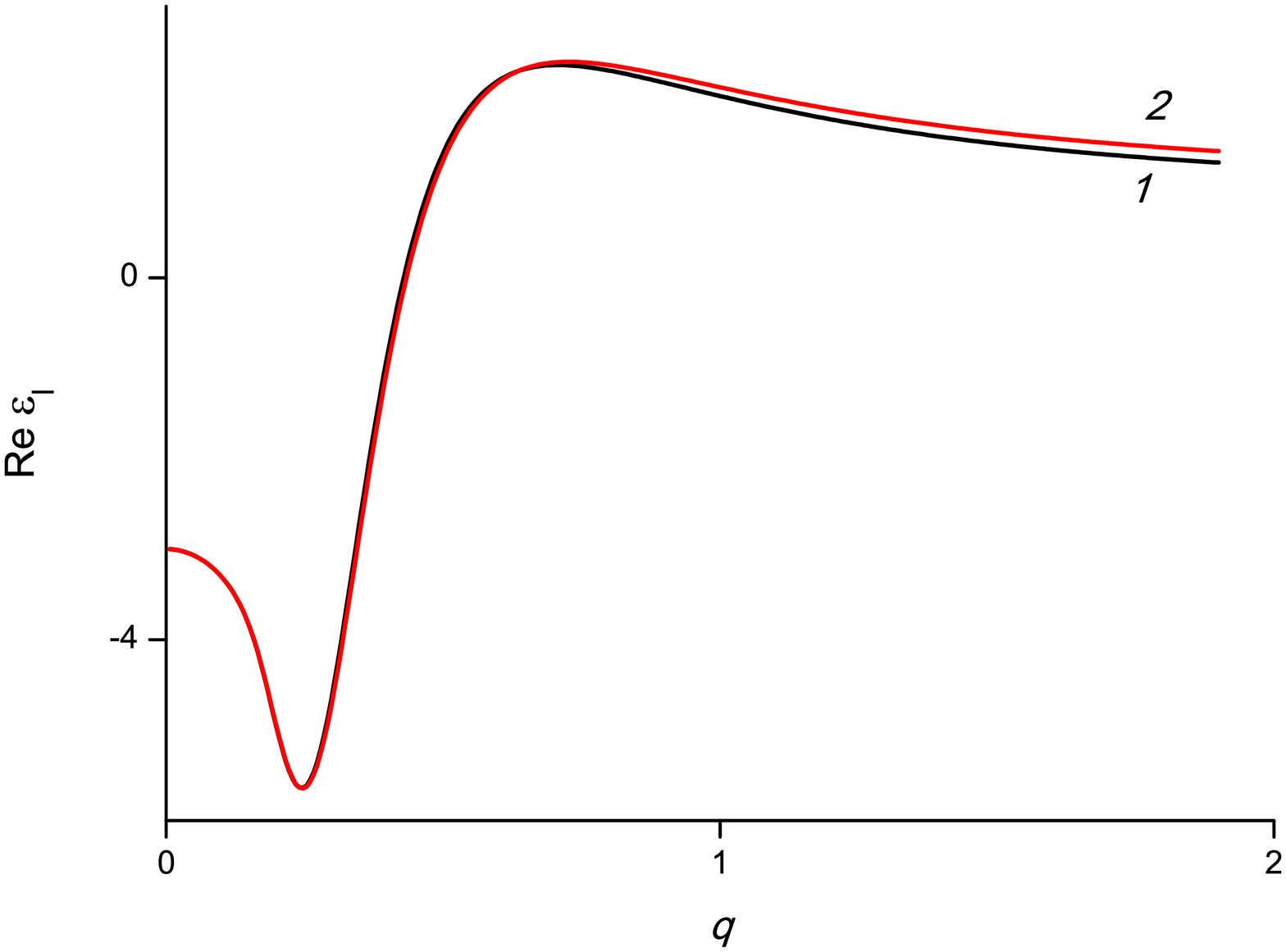}
\center{Fig. 1. Real parts of dielectric function, $x_p=1$, $x=0.5$,
$y=0.01$. }
\includegraphics[width=17.0cm, height=10cm]{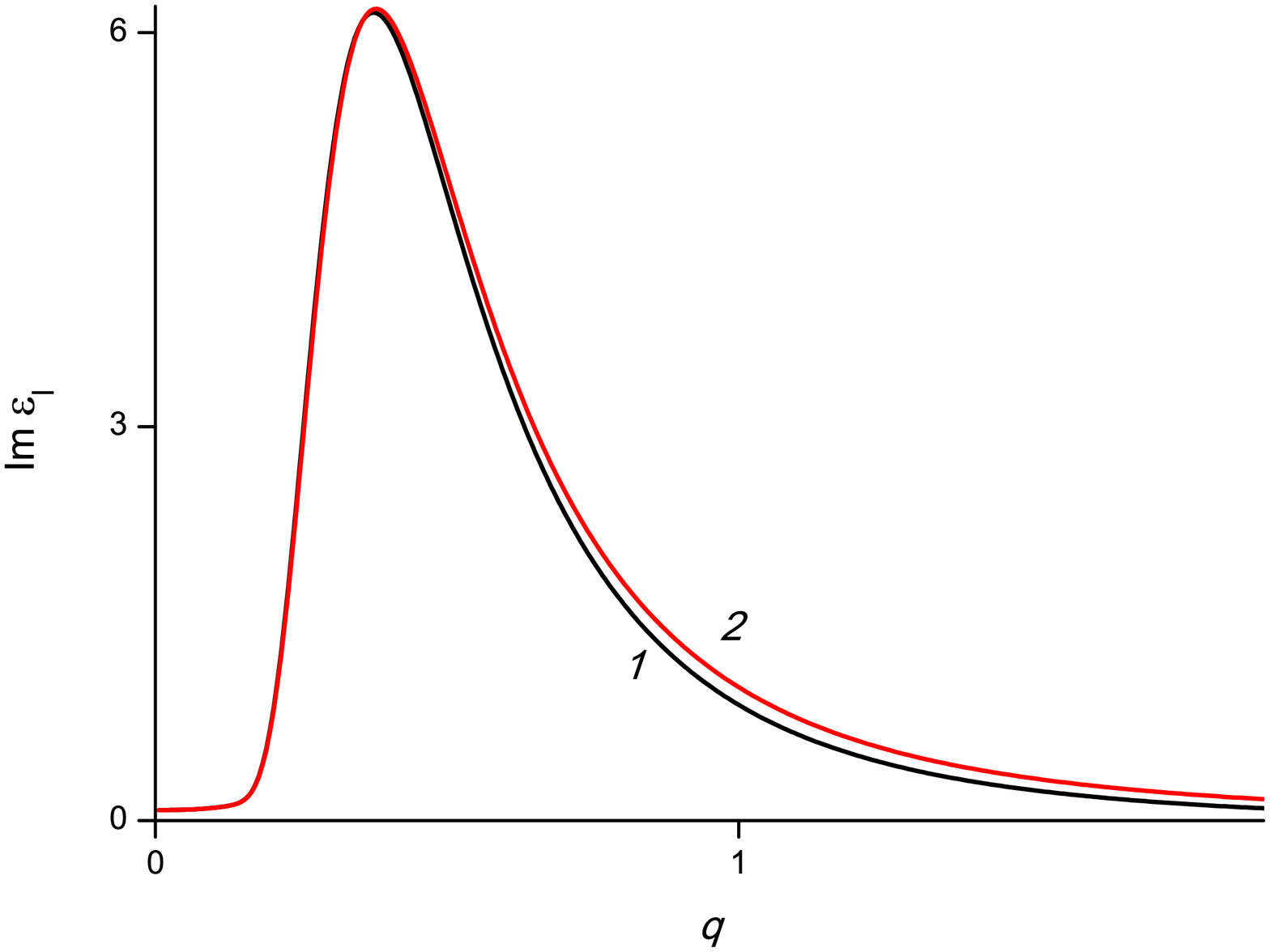}
\center{Fig. 2. Imaginary parts of dielectric function, $x_p=1$, $x=0.5$,
$y=0.01$. }
\end{figure}

\begin{figure}[ht]\center
\includegraphics[width=16.0cm, height=10cm]{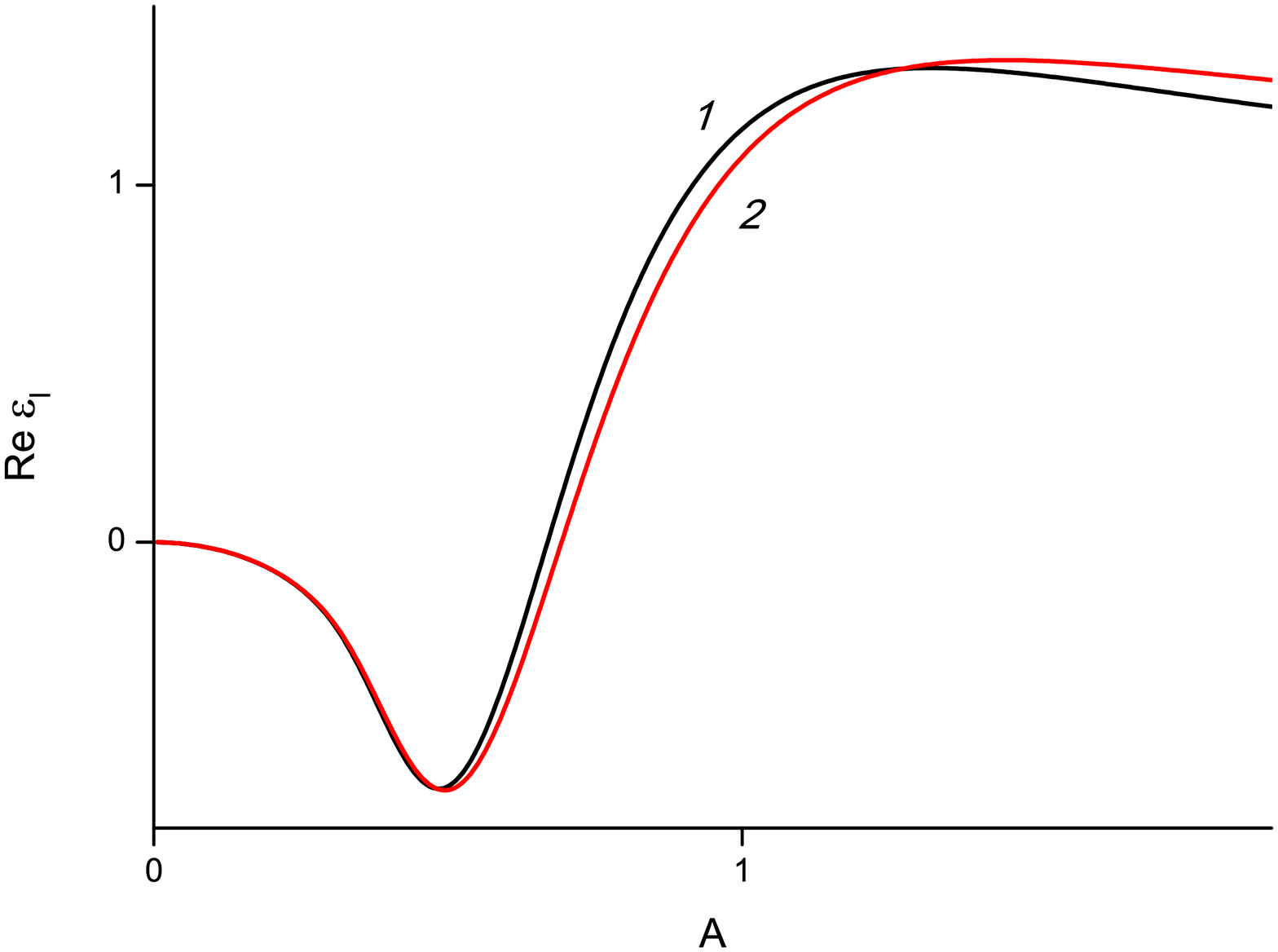}
\center{Fig. 3.  Real parts of dielectric function, $x_p=1$, $x=1$,
$y=0.01$. }
\includegraphics[width=17.0cm, height=10cm]{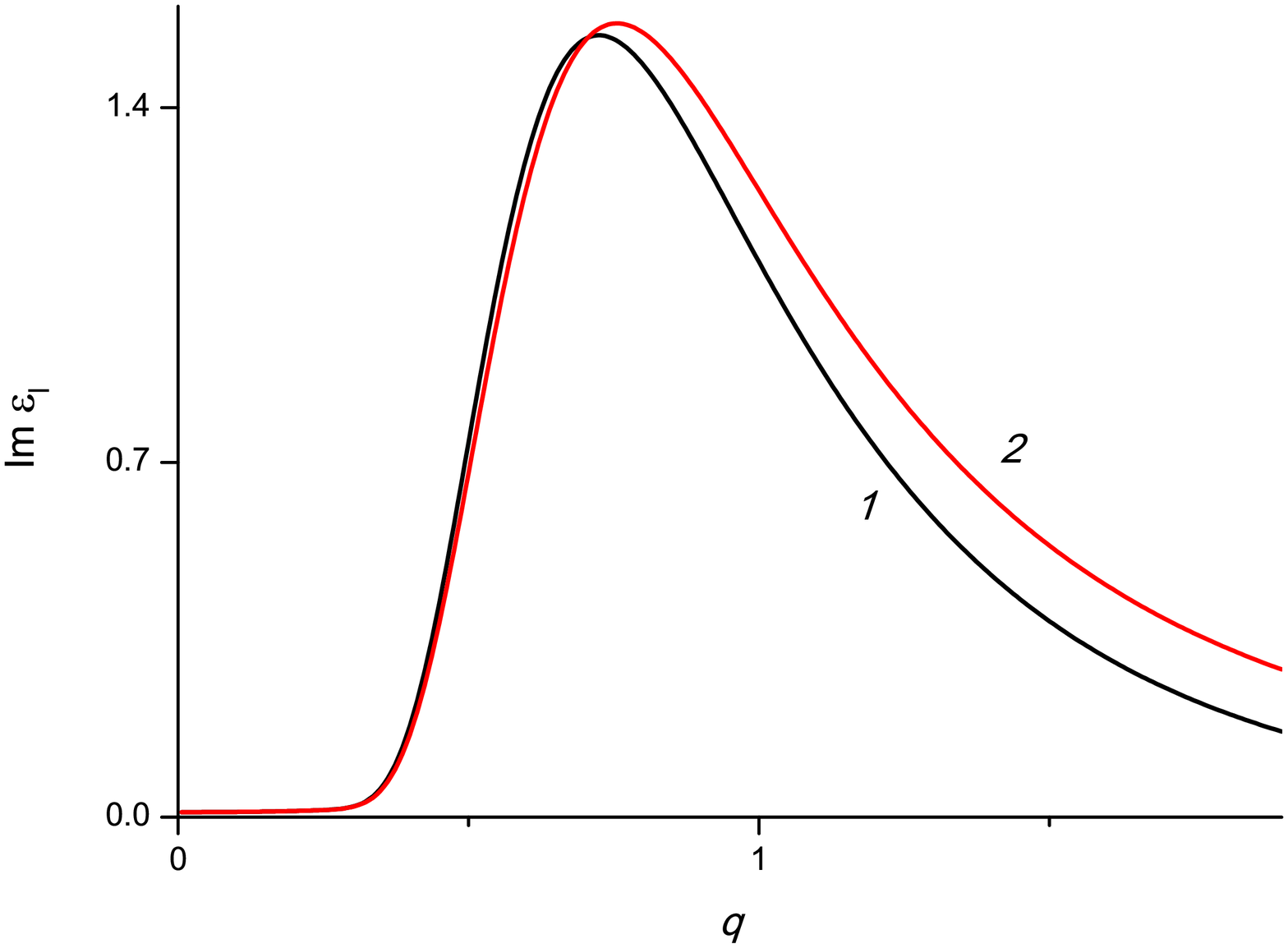}
\center{Fig. 4. Imaginary parts of dielectric function, $x_p=1$,$x=1$,
$y=0.01$.}
\end{figure}

\begin{figure}[ht]\center
\includegraphics[width=16.0cm, height=10cm]{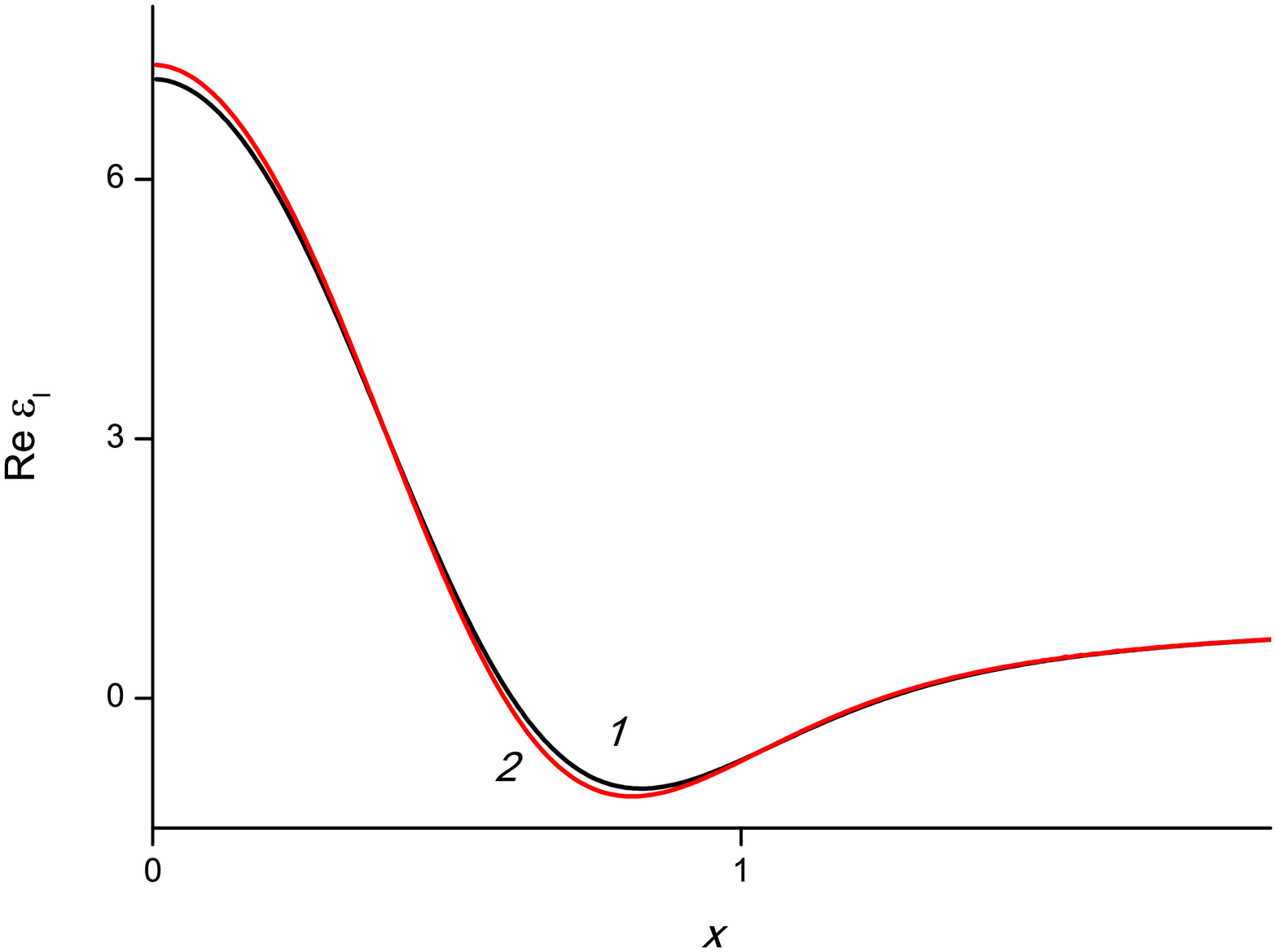}
\center{Fig. 5.  Real parts of dielectric function, $x_p=1$, $q=0.5$,
$y=0.001$.}
\includegraphics[width=17.0cm, height=10cm]{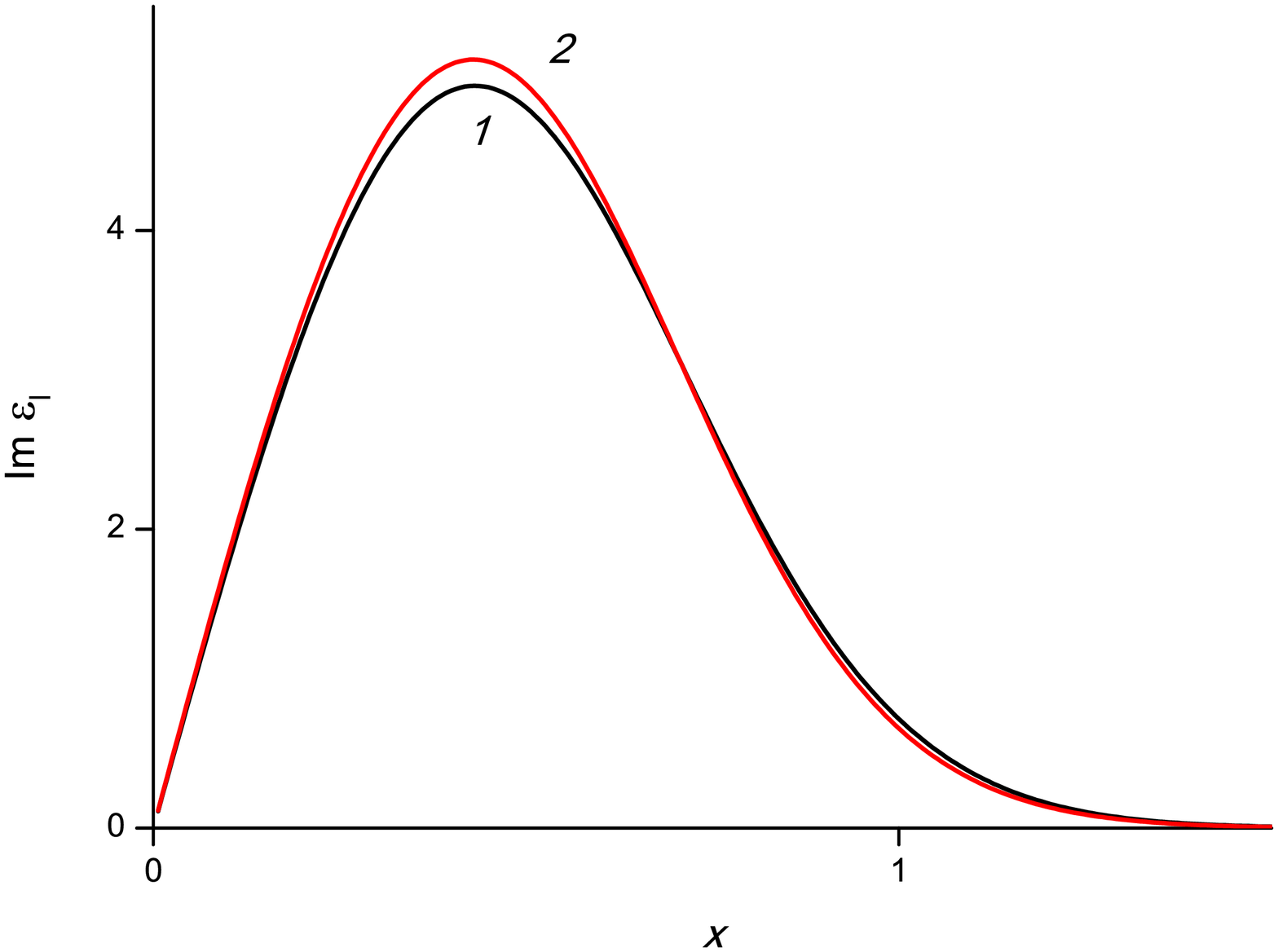}
\center{Fig. 6. Imaginary parts of dielectric function, $x_p=1$, $q=0.5$,
$y=0.001$. }
\end{figure}

\begin{figure}[ht]\center
\includegraphics[width=16.0cm, height=10cm]{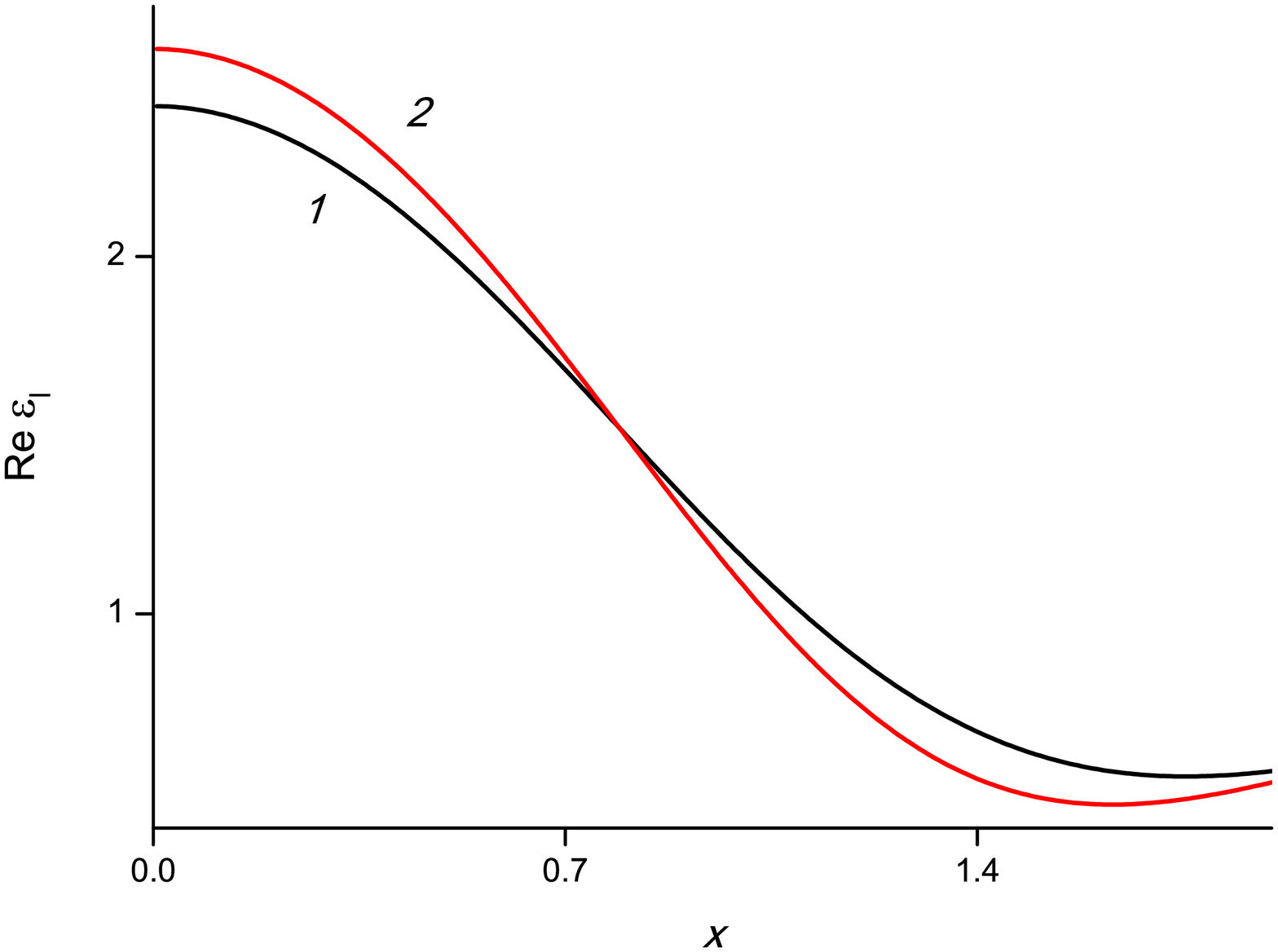}
\center{Fig. 7. Real parts of dielectric function, $x_p=1$, $q=1$,
$y=0.001$. }
\includegraphics[width=17.0cm, height=10cm]{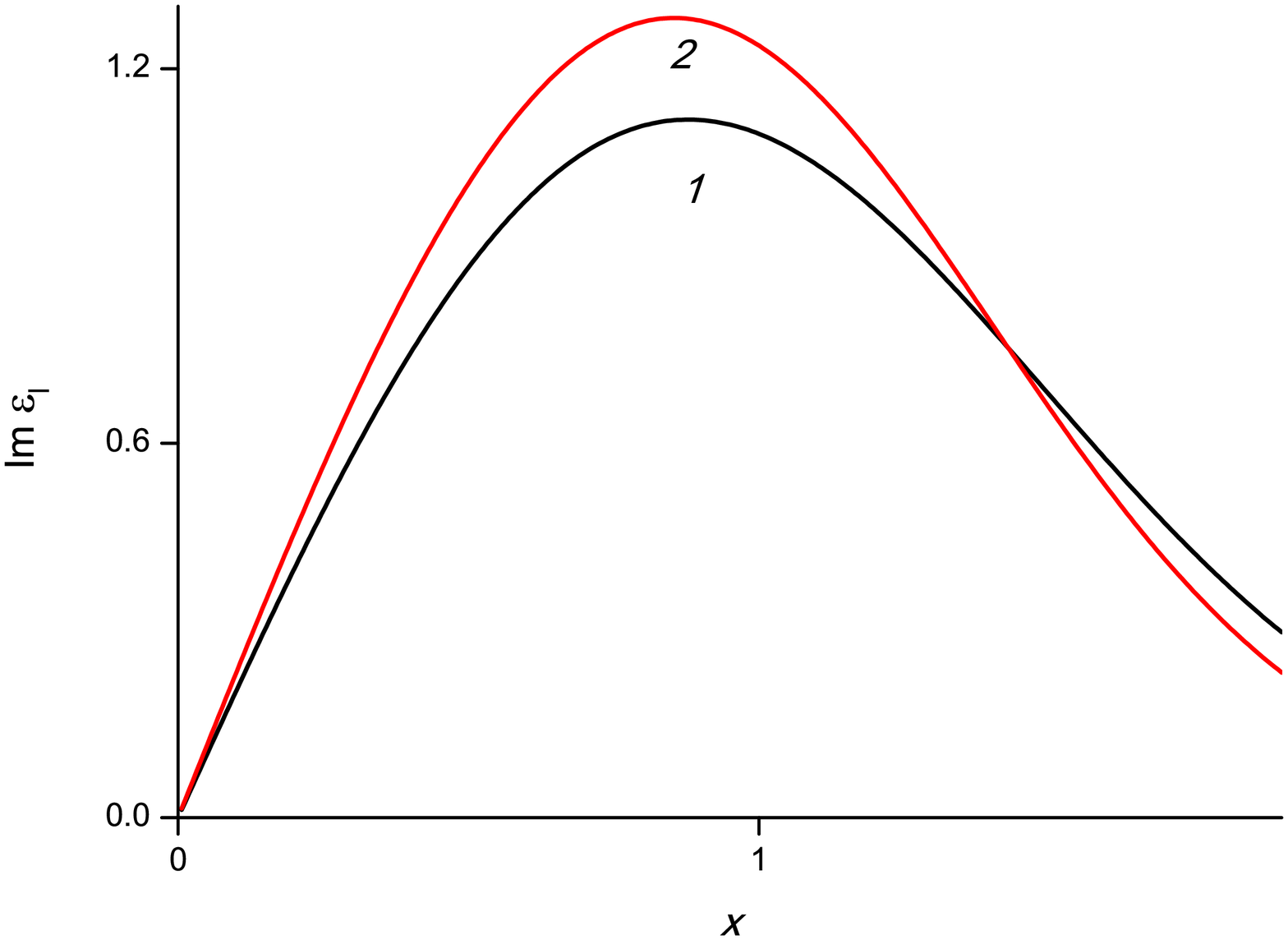}
\center{Fig. 8. Imaginary parts of dielectric function, $x_p=1$, $q=1$,
$y=0.001$.}
\end{figure}

\clearpage

\begin{center}
  \bf 5. Conclusion
\end{center}

The formula for dielectric function of non-degenerate and
maxwellian collisional plasmas is transformed to the form,
convenient for research.
Graphic comparison of longitudinal dielectric functions
of quantum and classical non-degenerate collisional plasmas is
made.

\end{document}